# A Gradient-based Deep Neural Network Model for Simulating Multiphase Flow in Porous Media

Bicheng Yan*, Dylan Robert Harp, Rajesh J. Pawar

Earth and Environmental Sciences, Los Alamos National Laboratory, Los Alamos, NM, USA, 87544

*Corresponding author

Email: bichengyan@lanl.gov (B. Yan); dharp@lanl.gov (D.R. Harp); rajesh@lanl.gov (R.J. Pawar)

**Abstract**

Simulation of multiphase flow in porous media is crucial for the effective management of subsurface energy and environment related activities. The numerical simulators used for modeling such processes rely on spatial and temporal discretization of the governing mass and energy balance partial-differential equations (PDEs) into algebraic systems via finite-difference/volume/element methods. These simulators usually require dedicated software development and maintenance, and suffer low efficiency from a runtime and memory standpoint for problems with multi-scale heterogeneity, coupled-physics processes or fluids with complex phase behavior. Therefore, developing cost-effective, data-driven models can become a practical choice since deep learning approaches are considered to be universal approximations. In this paper, we describe a gradient-based deep neural network (GDNN) constrained by the physics related to multiphase flow in porous media. We tackle the nonlinearity of flow in porous media induced by rock heterogeneity, fluid properties and fluid-rock interactions by decomposing the nonlinear PDEs into a dictionary of elementary differential operators. We use a combination of operators to handle rock spatial heterogeneity and fluid flow by advection. Since the augmented differential operators are inherently related to the physics of fluid flow, we treat them as first principles prior knowledge to regularize the GDNN training. We use the example of pressure management at geologic $CO_2$ storage sites, where $CO_2$ is injected in saline aquifers and brine is produced, and apply GDNN to construct a predictive model that is trained from physics-based simulation data and emulates the physics process. We demonstrate that GDNN can effectively predict the nonlinear patterns of subsurface responses including the temporal and spatial evolution of the pressure and $CO_2$ saturation plumes. GDNN has great potential to tackle challenging problems that are governed by highly nonlinear physics and enables development of data-driven models with higher fidelity.

**Keywords**: gradient-based deep neural network, partial differential equation, multiphase flow in porous media, carbon geological sequestration

## 1. Introduction

Many subsurface energy and environmental applications are controlled by multiphase fluid flow in porous media, including hydrocarbon recovery from oil and gas reservoirs [1, 8], geological $CO_2$ storage [6, 7, 26], geothermal energy extraction [14, 18, 21]. The management of such processes has been increasingly relying on high fidelity physics-based numerical simulations. The underlying physics are further complicated by heterogeneities across various length-scales [16, 46], complex fluid systems with multiple components and phases [8, 19], and coupled-physics processes with thermal transport and(or) geomechanical deformation [15, 29, 48]. Traditionally, the numerical simulators discretize the governing equations of multiphase flow in porous media by finite-difference/volume/element methods [1, 5, 8, 9, 12, 22, 32] resulting in large scale algebraic system of equations that may require significant computational resources and advanced numerical solvers and preconditioners [43, 44]. Recently, reduced-order methods (ROMs) have become popular alternatives for subsurface modeling. Particularly, dimensionality reduction techniques, such as multiscale methods [17, 20], Proper Orthogonal Decomposition (POD) [2] and the Krylov subspace projection method, are capable of identifying spatial-temporally coherent structures by reducing dimensionality. On the other hand, response surface models, such as kriging [51], polynomial regression and multivariate adaptive regression splines [49] are ROMs that can regress high dimensional data. These ROMs have been applied widely in solving flow in porous media achieving decent computational efficiency and accuracy [6, 10, 31, 38, 42, 50, 51].

During the past decade significant advances have been made in deep learning approaches [28] in both theory and application, which have exhibited great potential as universal approximators to map high dimensional data [3, 13, 24]. With recent advances in CPU and GPU-based parallel processing, deep learning has achieved great success in numerous applications including image recognition, natural language processing, and automated driving to name a few. However, most of these deep learning based tools are black-box solutions. In the scientific computing community, first principles with superior interpretability and transparency are the underlying rules governing traditional physics-driven modeling approaches. Similarly, some recent research using deep neural networks (DNN) leverages theoretical constraints to steer DNN training and enhance its predictive consistency with first principles. For example, Karpatne et al. [27] introduced physics-based terms into a DNN loss function, namely Physics-guided Neural Networks, and successfully applied it in lake temperature modeling. Raissi et al. [37] presented a Physics-Informed Neural Network (PINN), and leveraged the nonlinear PDE residuals to constrain the training of networks based on automatic differentiation [4, 33], and applied it to solve forward and inverse problems. In the area of fluid flow in porous media, Wang et al. [45] developed the Theory-guided Neural Network by imposing constraints based on practical engineering theories, physics principles and expert knowledge and used it to reliably predict subsurface single-phase flow in porous media. Fuks and Tchelepi [41] encoded the Buckley-Leverett displacement equation with two incompressible immiscible phases into the loss function of PINN, and investigated its predictive capability under different flux functions. Harp et al. [23] developed a physics-informed machine learning framework that couples a differentiable flow model, and validated its capability for subsurface pressure management in complex scenarios. The feasibility of approximating general PDE solutions by combining the PDEs with DNNs has been demonstrated to some extent by the above work. Nevertheless, the full evaluation of the nonlinear governing PDEs in many physics-based simulations across scales is quite expensive. Taking multiphase flow in heterogeneous porous media as an example, simulating fluid flow with complex phase behavior is notably nonlinear and numerically instable (convergence related) [8, 11, 19], or there might exist various constitutive relationships in different parts of the domain [30]. Repeatedly carrying the nonlinearity through the backpropagation of DNN training has significant computational costs and can introduce instability in the training process. Alleviating these limitations is extremely important to bring DNNs up to the scale of high-fidelity realistic simulations.

The focus of this work is to explore the use of DNN as a universal approximator to learn the multiphase flow in homogeneous and heterogeneous porous media, aiming to build a DNN-based modeling framework as a computationally efficient alternative to physics-based reservoir simulators. Given the complexity of multiphase flow in heterogeneous porous media, we do not attempt to regularize the training loss function of the neural network via the closed-form residuals of nonlinear PDEs. Instead, our main contribution is to transform the PDEs into a dictionary of elementary differential operators or gradients that are fundamental units of the PDEs, namely a Gradient-based DNN (GDNN). As the differential operators inherently represent patterns associated with physics of fluid flow, e.g., the fluid flow velocity field, we treat them as first principles prior knowledge to steer the GDNN training. For example, since pressure plume patterns in heterogeneous porous media are dominated by advective flux, a second order differential operator combining spatial gradients of pressure and permeability is proposed to constrain the phenomenon. The prediction of these operators is computed by automatic differentiation during the training, and their ground truth from reservoir simulation data is calculated by a central finite-difference scheme prior to the training. Therefore, the loss function consists of the mismatch of state variables (e.g., pressure and saturation) and the newly proposed mismatch of the elementary differential operators.

The rest of the paper is organized as follows. In Section 2, we describe the governing equation of multiphase flow in porous media and introduce the basic formulation of the GDNN modeling framework. In Section 3, we use a 2D homogeneous example to clearly demonstrate our GDNN approach. Then, we switch to a 3D heterogeneous example to demonstrate GDNN's improved ability to capture irregular pressure and saturation patterns compared to a DNN. Finally, we conclude our findings in Section 4.

## 2. Governing Equation of Multiphase Flow in Porous Media and Formulation of GDNN

In this section we first describe the governing equation of multiphase flow in porous media, and illustrate the formulation of the GDNN model to predict pressure and saturation.

### 2.1 Governing Equation of Multiphase Flow in Porous Media

The primary fluid phases in a saline $CO_2$ storage reservoir include aqueous and supercritical phases. Water is the major component of the aqueous phase while $CO_2$ makes up majority of the supercritical phase. The solubility of $CO_2$ in the aqueous phase and the vaporization of water in the supercritical phase is taken into account in our physics simulation data [11] used for training the deep learning models, even though both are not significant. For the purpose of clear illustration of the mathematics and easier extraction of differential operators for GDNN, here the phenomenon of $CO_2$ dissolution and water vaporization is omitted in the following formulation. Therefore, the problem can be mathematically formulated as a general multiphase flow problem, and the mass balance governing equation for each phase is,

$$\frac{\partial(\phi \rho_\alpha S_\alpha)}{\partial t} - \nabla \cdot \left( \frac{K k_{r\alpha} \rho_\alpha}{\mu_\alpha} (\nabla p_\alpha - \rho_\alpha g \nabla z) \right) + q_\alpha = 0 \qquad (1)$$

where the first term is the accumulation term for fluid storage within the rock pore volume; the second is the flux due to advection; the third is the source or sink term; subscript $\alpha = w, g$ represents the aqueous and supercritical phases, respectively; $\rho_\alpha$ is the fluid phase density at the reservoir condition; $S_\alpha$ is the phase saturation; $\phi$ is the rock porosity; $K$ is the rock intrinsic permeability; $k_{r\alpha}$ is the phase relative permeability; $\mu_\alpha$ is the fluid phase viscosity; $p_\alpha$ is the fluid phase pressure; $g$ is the acceleration due to gravity; $z$ is depth; $q_\alpha$ is the source/sink rate for extracting or injecting fluid into the reservoir usually calculated through the Peaceman well model for numerical simulations [35].

Additionally, several auxiliary relationships constrain the variables in the mass balance equations. The pore space is filled by both fluid phases, shown as,

$$\sum S_\alpha = 1 \qquad (2)$$

The capillary pressure relates the phase pressures between wetting and non-wetting phases. In a saline aquifer the wetting phase is typically the aqueous phase, and the non-wetting one is the supercritical phase, shown as,

$$p_c = p_g - p_w \qquad (3)$$

Further, fluid viscosity $\mu_\alpha$ and density $\rho_\alpha$ are usually nonlinear functions of pressure and temperature, and phase relative permeability is a function of fluid phase saturation $S_\alpha$. Since the supercritical phase of $CO_2$ is much more compressible than its aqueous phase, the supercritical phase properties have strong nonlinear relationships with pressure. Typically, density and viscosity of supercritical fluid are computed using an Equation of State (EOS), such as the Peng-Robinson EOS [36] which relates fluid properties to pressure and temperature.

## 2.2 Formulation of GDNN to Predict Pressure and Saturation

In this sub-session, we start with a discussion of physics-informed neural networks (PINNs) [37] to briefly provide the background about using neural networks to solve general partial-differential equations (PDEs). Motivated by that, we propose the Gradient-based Deep Neural Network (GDNN) to tackle highly nonlinear PDEs in problems such as multiphase flow in porous media. The major difference between PINN and GDNN is that PINN approximates general PDEs by directly imposing them on the training loss function, whereas GDNN transform highly nonlinear PDEs into essential elementary differential operators that are applied to training loss function.

The PDEs governing the multiphase flow in porous media are typically written as conservation laws [37] in the general form shown as,

$$f = u_t + \mathbb{N}[u] = 0, x \in \Omega, t \in [0,T] \tag{4}$$

where $f$ is the residual of the PDE, $u = u(t,x)$ is the state variable, $u_t$ represents $\frac{\partial u}{\partial t}$, and $\mathbb{N}[\cdot]$ is a combination of multiple nonlinear differential operators.

In the PINN approach, a deep neural network (DNN), specifically a deep feedforward network, is used as an approximator of nonlinear PDEs. The DNN is usually assembled as a layered chain structure, and each layer is a function of the layer preceding it, shown as,

$$F_l = \sigma(W_l^T F_{l-1} + b_l), where\ l = 1, 2, \cdots L \tag{5}$$

where $F_l$ is the output of layer $l$, $\sigma$ denotes a nonlinear activation function, $W_l$ and $b_l$ represents the weight matrix and bias term of layer $l$, respectively. The layer sequence starts with input layer $F_0$ and ends with output layer $F_L$, and all other layers are referred to as hidden layers.

To approximate the residual $f$ in Equation (4), Raissi et al. [37] modified the output layer of the PINN after obtaining an approximation of the state variable $u$ denoted $\hat{u}$. With the aid of automatic differentiation [4], $\hat{u}$ is further differentiated to construct $\hat{f}$ that approximates $f$ in Equation (4). Since the true solution of $f$ is 0, $|f - \hat{f}| = |\hat{f}|$ is treated as a regularization term in the loss function during PINN training. The training process of PINN minimizes the loss function $\mathcal{L}$, which has two components, a data-driven mismatch component and a physics-driven PDE residual component, shown as,

$$\mathcal{L} = |u - \hat{u}| + |f - \hat{f}| = |u - \hat{u}| + |\hat{f}| \tag{6}$$

In this work, GDNN is tasked with emulating the predictive process of a physics-based multiphase flow simulator, and thus it is required to learn the patterns generated by applying the mass balance equation (Equation (1)) into multiphase flow in porous media. The mass balance equation belongs to the generalized format of Equation (4) where pressure and phase saturation are the state variables $u$. When the fluids are only slightly compressible or incompressible, the differential operators contained in $\mathbb{N}[\cdot]$ in the mass balance equation are relatively straightforward to construct and converge in neural network models [41]. However, in this work we have a highly compressible fluid, the supercritical phase of $CO_2$, which has complex phase behavior [19] with properties varying nonlinearly with pressure and temperature [47], and this leads to higher computational cost for high-fidelity numerical simulations. For example, in a fully compositional reservoir simulation, solution of an EOS model to calculate fluid properties can take up to 70% of the total computational time [43]. Therefore, it is not computationally efficient to differentiate an extremely expensive EOS model in order to obtain a closed-form solution of physics-driven PDE loss in Equation (6), and repeatedly solve it during each training epoch over all training samples.

Unlike the PINN approach [37], we do not directly impose the PDEs on the loss function during the training process of GDNN. Instead, we decompose the highly nonlinear PDE $f$ to the level of the elementary differential operators $\{u_t, u_x, u_{xx}, \cdots\}$, which capture essential parts of the physics, e.g., $\nabla p$ in the mass balance Equation (1) controls the fluid advective flux based on the Darcy's law. Therefore, the original closed-form PDE loss $|f - \hat{f}|$ in Equation (6) is transformed into a series of elementary differential loss sub-terms, which provide physical constraints based on first principles inherent in the reservoir simulation results, shown as,

$$|f - \hat{f}| \longrightarrow \sum |D - \hat{D}|, \text{with } D \in \{u_t, u_x, u_{xx}, \cdots\} \text{ and } \hat{D} \in \{\hat{u}_t, \hat{u}_x, \hat{u}_{xx}, \cdots\} \tag{7}$$

Substituting the transformation in Equation (7) into Equation (6), the loss $\mathcal{L}$ of the GDNN becomes,

$$\mathcal{L} = |u - \hat{u}| + \sum |D - \hat{D}|. \tag{8}$$

Therefore, in Equation (8) we do not directly evaluate the close-form solution of PDEs. This is extremely favorable when the terms in the conservation laws are highly nonlinear due to nonlinear constitutive relations, e.g., the density and viscosity of supercritical phase in Equation (1), or when we are actually not clear about the close-form PDE residual except several essential elementary differential operators. Additionally, it removes the need to assume some reduced form of physics for the purpose of proposing a closed-form of loss for the conservation laws.

In multiphase flow simulations in porous media, the state variables to predict are the pressure and phase saturations. The solution of these state variables varies with spatial $(x, y, z)$ and time $(t)$ coordinates, and is usually controlled by multiple parameters, such as permeability $(K)$ and porosity $(\phi)$, fluid thermodynamic properties, rock-fluid interaction and well control, etc. In geologic $CO_2$ storage operations, $CO_2$ injection rate $q_{inj}$ is an operational parameter that may be varied over time. In certain cases, water may be produced to manage reservoir pressure in order to reduce leakage and induced seismicity risks. In our current work, we have considered the scenario of pressure management at a geologic $CO_2$ storage operation where the aim is to manage reservoir pressure through water production. Typically, only a few wells are used for $CO_2$ injection and water production and they are sparsely located within the study domain. Therefore, it is necessary to inform the GDNN of locations that are producers ($\sim producer$) or injectors ($\sim injector$). In the following, a feature list is defined as the input variables of the GDNN model to predict the state variables. In homogeneous permeability fields, the spatial derivatives of permeability are constant zeros, so they are not needed in the feature list. Therefore, for problems with homogeneous permeability fields, the features list to predict pressure and saturation is shown as,

$$[p, S_g]_\Omega = f(x, y, t, \phi, K, q_{inj}, \sim producer, \sim injector) \tag{9}$$

On the other hand, for problems with heterogeneous permeability fields, the features list includes the spatial derivatives of permeability for capturing the effect of its spatial variability, which is shown as,

$$[p, S_g]_\Omega = f(x, y, z, t, \phi, K, q_{inj}, \sim producer, \sim injector, \frac{dK}{dx}, \frac{dK}{dy}, \frac{dK}{dz}) \tag{10}$$

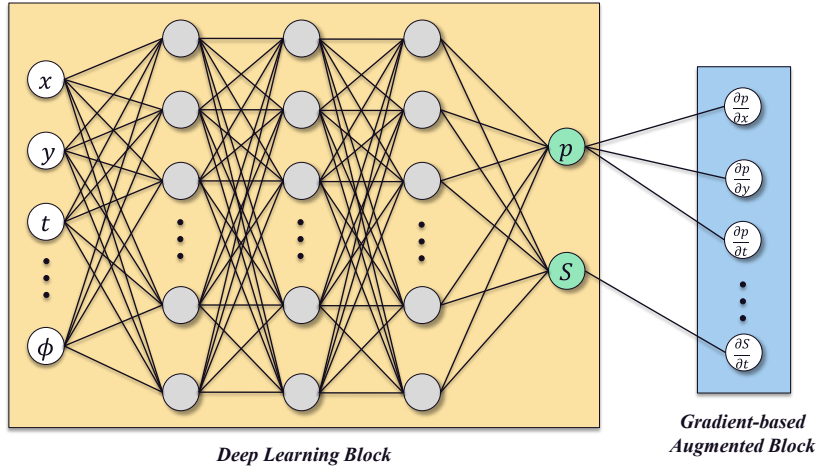

Fig. 1. Functional blocks in Gradient-based Neural Network Model

The various elements of a GDNN used for our problem are shown in **Fig. 1**. It includes two functional blocks, the deep learning block (DNN) using PyTorch functionality [34], and the gradient-based augmented block to compute the elementary differential operators by automatic differentiation [33]. The gradient-based augmented block is synchronized with the deep learning block during the backpropagation procedure.

The fluid accumulation term in the mass balance equation (Equation (1)) can be transformed to link to the pressure and saturation derivatives with regards to time via the chain rule,

$$\frac{\partial(\phi \rho_\alpha S_\alpha)}{\partial t} = \phi \rho_\alpha \frac{\partial S_\alpha}{\partial t} + S_\alpha \left[\rho_\alpha \frac{\partial \phi}{\partial t} + \phi \frac{\partial \rho_\alpha}{\partial t}\right] = \phi \rho_\alpha \frac{\partial S_\alpha}{\partial t} + S_\alpha \left[\rho_\alpha \frac{\partial \phi}{\partial p} + \phi \frac{\partial \rho_\alpha}{\partial p}\right] \frac{\partial p}{\partial t} \quad (11)$$

Therefore, the fluid accumulation term is related to the first order temporal gradients of pressure and saturation. To capture the physics embedded in the training data, the following differential operators are computed in the augmented block:

The first order spatial differentials of pressure, $\nabla p = [\frac{\partial p}{\partial x}, \frac{\partial p}{\partial y}, \frac{\partial p}{\partial z}]$: from the flux term $\nabla \cdot \left(\frac{K k_{r\alpha} \rho_\alpha}{\mu_\alpha}(\nabla p_\alpha - \rho_\alpha g \nabla z)\right)$ in Equation (1) and applied to the scenario with homogeneous permeability;

The second order spatial differentials of pressure, $\nabla \cdot J \cong \frac{\partial}{\partial x}\left(K_x \frac{\partial p}{\partial x}\right) + \frac{\partial}{\partial y}\left(K_y \frac{\partial p}{\partial y}\right) + \frac{\partial}{\partial z}\left(K_z \frac{\partial p}{\partial z}\right)$: related to the flux term $\nabla \cdot \left(\frac{K k_{r\alpha} \rho_\alpha}{\mu_\alpha}(\nabla p_\alpha - \rho_\alpha g \nabla z)\right)$ in Equation (1) and applied to the scenario with heterogeneous permeability;

The first order time differential of pressure, $\frac{\partial p}{\partial t}$: related to the accumulation term $\frac{\partial(\phi \rho_\alpha S_\alpha)}{\partial t}$ by Equation (11);

The first order time differential of saturation, $\frac{\partial S_\alpha}{\partial t}$: related to the accumulation term $\frac{\partial(\phi \rho_\alpha S_\alpha)}{\partial t}$ by Equation (11).

The ground truth for the differential operators are computed from numerical simulation output by using the central finite difference scheme prior to training the GDNN. For example, with the pressure field $p$ predicted by high-fidelity numerical simulations, $\frac{\partial p}{\partial x}$ and $\frac{\partial p}{\partial y}$ can be approximated across the model domain. Example pressure-gradient fields for the homogeneous permeability scenario are shown in **Figs. 2(b)** and **(c)**. The GDNN learns patterns from these gradients which embed the physics associated with the fluid velocity field into the neural network.

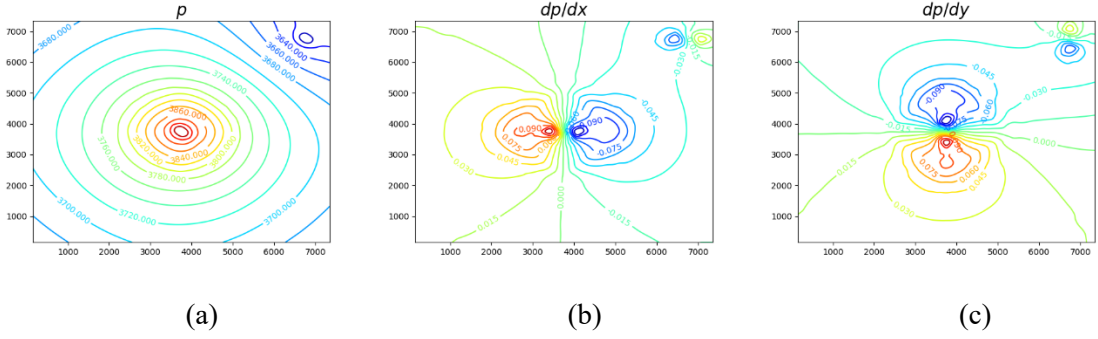

(a)            (b)            (c)

Fig. 2. Example pressure field and corresponding spatial gradients based on central finite difference: (a) $p$; (b) $\frac{\partial p}{\partial x}$; (c) $\frac{\partial p}{\partial y}$.

We treat the DNN (represented by the Deep Learning Block in **Fig. 1**) as a baseline in our analysis to gauge the relative performance of the GDNN. As shown in **Table 1**, three options for loss function are derived using the analysis above. In the table, the operator $\frac{1}{N}\sum|\cdot|^2$ is the mean square error (MSE) where $N$ is the number of samples. The loss function of the DNN is the mismatch in state variables (pressure and saturation). On the other hand, there are two options for the loss function of the GDNN, depending on whether the permeability field is homogeneous or heterogeneous. In the homogeneous scenario, the loss function includes the first order spatial and temporal gradients of state variables. In the heterogeneous scenario, the first order pressure spatial gradient terms are replaced with second order flux gradient that include both permeability and pressure. DNN and GDNN in this work were trained on a single GPU (NVIDIA Quadro RTX 4000).

Table 1. Loss function options for deep learning.

| Options | Loss Function | Scenario |
|---|---|---|
| DNN | $\mathcal{L} = \frac{1}{N_p}\sum|p-\hat{p}|^2 + \frac{1}{N_s}\sum|S-\hat{S}|^2$ | All cases |
| GDNN-Option 1 | $\mathcal{L} = \frac{1}{N_p}\sum|p-\hat{p}|^2 + \frac{1}{N_s}\sum|S-\hat{S}|^2 + \frac{1}{N_c}(\sum|p_x-\hat{p}_x|^2 + \sum|p_y-\hat{p}_y|^2 + \sum|p_z-\hat{p}_z|^2 + \sum|p_t-\hat{p}_t|^2 + \sum|S_t-\hat{S}_t|^2)$ | Homogeneous permeability field |
| GDNN-Option 2 | $\mathcal{L} = \frac{1}{N_p}\sum|p-\hat{p}|^2 + \frac{1}{N_s}\sum|S-\hat{S}|^2 + \frac{1}{N_c}(\sum|\nabla\cdot J - \nabla\cdot\hat{J}|^2 + \sum|p_t-\hat{p}_t|^2 + \sum|S_t-\hat{S}_t|^2)$ | Heterogeneous permeability field |
| Here $f_x = \frac{\partial f}{\partial x}$, and $\hat{f}$ is the prediction from machine learning, where $f$ is either $p$, $S$, or $J$. | | |

## 3. Results

### 3.1 Homogenous Two Dimensional Two-Phase Flow

In this section, we use a 2D homogeneous reservoir model with a grid composed of 25 by 25 cells in the $x$ and $y$ directions. The grid dimension is $300\,ft$ by $300\,ft$ by $33.3\,ft$ in the $x, y$ and $z$ directions, respectively. As shown in **Fig. 3**, there is a $CO_2$ injection well in the center of the reservoir at grid cell (13, 13), and a water production well in the lower right corner at grid cell (23, 23). The final simulation time is 5.9 years with 72 time steps. The permeability and porosity fields are homogeneous, and the $CO_2$ injection well is constrained by a fixed injection rate, and the water production well is set to maintain a constant bottom-hole pressure of 3525 psia to manage reservoir pressure. The simulations were performed using a commercial reservoir simulator CMG GEM [11]. In total 27 simulations were performed with multiple combinations of permeability, porosity and $CO_2$ injection rate.

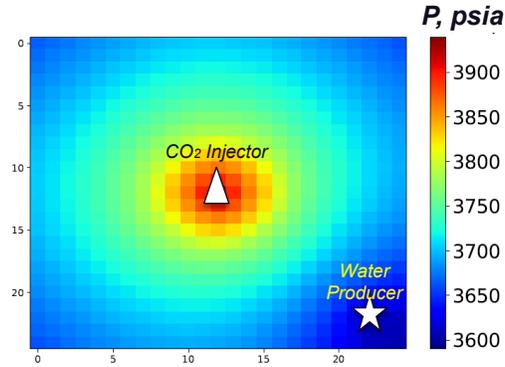

Fig. 3. Reservoir model with a pair of $CO_2$ injection and water production wells

The relative permeability curves of supercritical and aqueous phases are shown in **Fig. 4**. The permeability and porosity fields are correlated with a semi-log linear relationship shown by the cross-plot in **Fig. 5(a)**. The cross-plot of permeability and $CO_2$ injection rate is shown in **Fig. 5(b)** along with indications of which cases were used for training (24 cases, 90% of samples) and for testing (3 cases, 10% of samples).

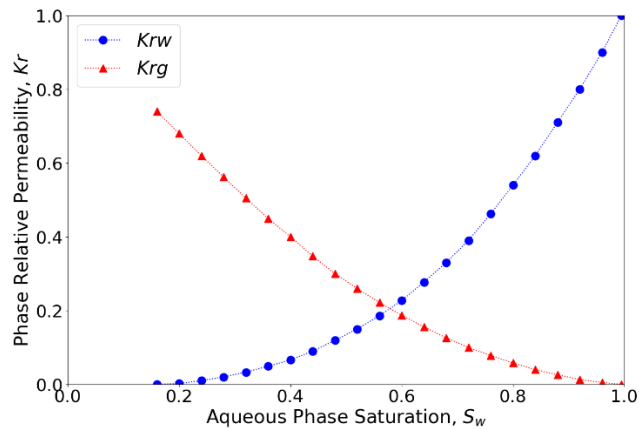

Fig. 4. Relative permeability curves for supercritical ($g$) and aqueous ($w$) phases

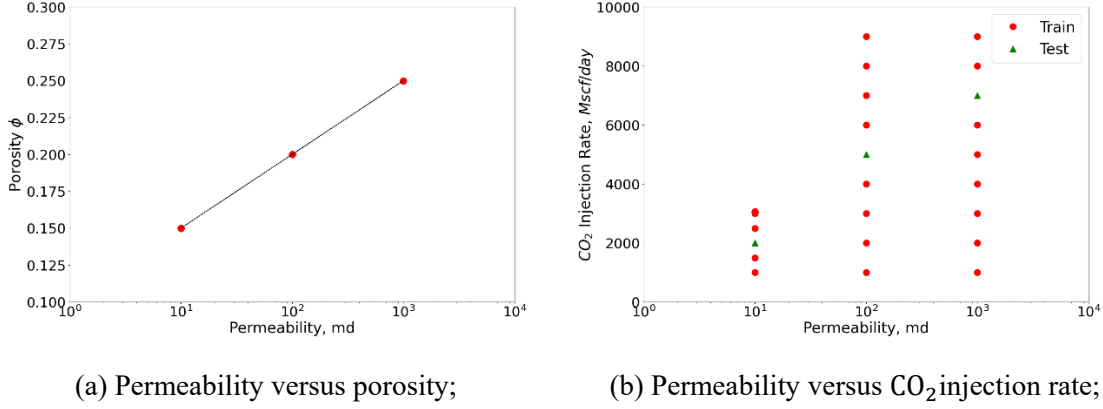

(a) Permeability versus porosity;  (b) Permeability versus $CO_2$ injection rate;

Fig. 5. The cross-plot of sampling variables

Both, DNN and GDNN-Option 1 (**Table 1**) are trained to learn the underlying physics of multiphase flow for this 2D homogeneous problem. There are 1,065,000 training data samples in a single batch. The training process used the LBFGS optimizer in PyTorch [34]. The deep learning block architecture (**Fig. 1**) includes 8 input variables in the input layer (refer to Equation (9)), 8 hidden layers with 20 neurons per hidden layer, 2 output variables $[p, S_g]$ in the output layer, and the activation function, $Tanh(x) = \frac{e^x - e^{-x}}{e^x + e^{-x}}$. Given that the gradient contours (**Fig. 2**) are relatively smooth, only 5% of training data points are selected for GDNN-Option 1 model and it inherits the weights and bias values from the DNN model as a starting point for training (transfer learning). However, the source and sink data points (boundary conditions) are always included in the GDNN training to guarantee the resolution at the sources and sinks. The training time was 20 minutes for DNN and 30 minutes for GDNN-Option 1. The increased training time for GDNN is primarily due to computations of the gradients based on automatic differentiation. In order to evaluate the stability of temporal predictions of these models, the temporal error is calculated as,

$$Y_{error}^t = \frac{1}{n_c n_g} \sum_{i=1}^{n_c} \sum_{j=1}^{n_g} \frac{\left\| Y_{i,j}^t - \hat{Y}_{i,j}^t \right\|}{\left(Y_{i,j}^t\right)_{max} - \left(Y_{i,j}^t\right)_{min}} \quad (12)$$

which is modified from the temporal error equation in Tang et al. [40], where $t$ is the time step; $Y$ is the pressure $p$ or the supercritical phase saturation $S_g$; $n_c$ is the number of testing simulation runs; $n_g$ is the number of grid cells of each reservoir simulation run.

In **Fig.6**, we present the temporal error propagation of pressure and supercritical phase saturation of all testing cases for different models. Hereafter we shorten "supercritical phase saturation" to "saturation". Both DNN and GDNN-Option 1 models exhibit very low pressure temporal error (<1%), and saturation error (<0.1%). The pressure temporal error slightly decreases over time, while the saturation temporal error slightly increases with time, which might be caused by the architecture of the network that simultaneously predicts pressure and saturation in DNN and GDNN-Option 1. Further, the small temporal errors of the prediction give us good confidence in prediction of pressure and saturation for the whole life of the reservoir. The overall performance of GDNN-Option 1 is slightly better than DNN, and the advantage of GDNN-Option 1 will be further demonstrated by the results of individual testing cases below.

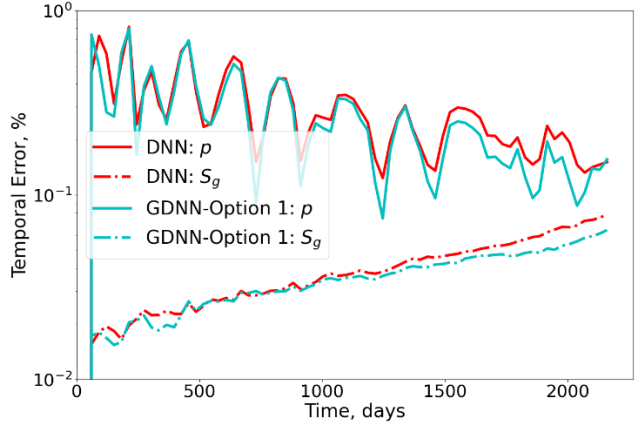

Fig. 6. Temporal error propagation for pressure and saturation in homogeneous scenario

Next, we discuss the predictions of pressure and saturation of 3 testing cases. **Table 2** lists the parameter values associated with the 3 cases. The testing cases include low, medium and high values from the parameter ranges to ensure that there is no sampling imbalance in the training-testing case split.

Table 2. Value of parameters for 3 different testing cases

| Case | Permeability, md | Porosity | $CO_2$ injection rate, $mscf/day$* |
|---|---|---|---|
| 4 | 10 | 0.15 | 2000 |
| 14 | 100 | 0.20 | 5000 |
| 25 | 1000 | 0.25 | 7000 |
| Overall Range for 27 Cases | 10 to 1000 | 0.15 to 0.25 | 1000 to 9000 |

*$mscf/day$: 1000 standard cubic feet per day

Two metrics are used to evaluate the accuracy of the predicted pressure and saturation snapshots for individual testing cases, including mean absolute error ($e_1$) and mean relative error ($e_2$) shown as,

$$e_1 = \frac{1}{n_g}\sum_{j=1}^{n_g}\|Y_j - \hat{Y}_j\| \tag{13}$$

$$e_2 = \frac{1}{n_g}\sum_{j=1}^{n_g}\frac{\|Y_j-\hat{Y}_j\|}{Y_j} \times 100\% \tag{14}$$

Where $Y$ is the pressure $p$ or saturation $S_g$; $n_g$ is the number of grid cells in a single $Y$ snapshot.

Since we have already discussed the stability of temporal error propagation, only the pressure and saturation snapshots at 5.9 years are presented in **Fig. 7** through **9**. For Case 4 and Case 14 (**Fig. 7(a)** and **8(a)**), the pressure plume prediction of DNN is quite smooth, and it captures the global contours well. However, the mean absolute error between DNN and reservoir simulations is relatively high at the well locations compared to the rest of the domain. On the other hand for GDNN-Option 1 (**Fig. 7(b)** and **8(b)**) the pressure errors at the water production well location are significantly lower than that in DNN, and the error distribution is more uniform than that in DNN. For Case 25, DNN does not perform as well for predicting pressure as the other two test cases (**Fig. 9(a)**), as reflected in the discontinuous pressure plume which is physically infeasible. On the contrary, GDNN-Option 1 predicts a continuous pressure plume (**Fig. 9(b)**) because the pressure gradients used in the training loss ensure smooth pressure plumes which effectively reflect those observed from the ground truth based on reservoir simulation output. For all the 3 cases, both DNN and GDNN-Option 1 predict the $CO_2$ plume shape with high accuracy (**Figs. 7(c)**, **7(d)**, **8(c)**, **8(d)**, **9(c)** and **9(d)**). The better prediction of saturation compared to pressure is due to the fact that the saturation plume is significantly smaller and less complex than the pressure plume and localized around the $CO_2$ injection well. The aggregated mean relative errors of pressure over all the time snapshots for the 3 cases are 0.29% for DNN and 0.26% for GDNN-Option 1, and similarly the aggregated errors of saturation are 0.039% for DNN and 0.035% for GDNN-Option 1. The results demonstrate that for homogenous permeability and porosity fields, GDNN-Option 1 can predict pressure and saturation plumes more accurately than DNN and can ensure smoothness in the emulated pressure field.

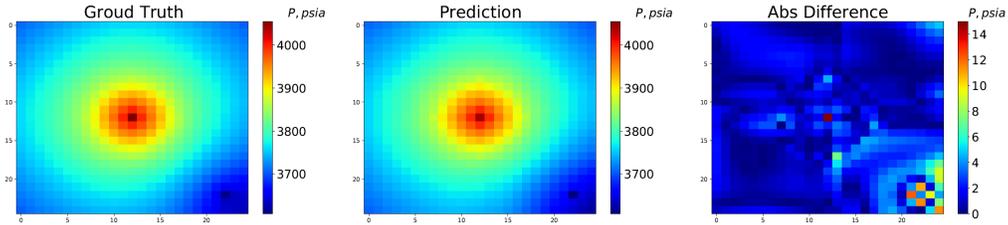

(a) Pressure from DNN: mean absolute error = 1.469 psia, mean relative error = 0.328%

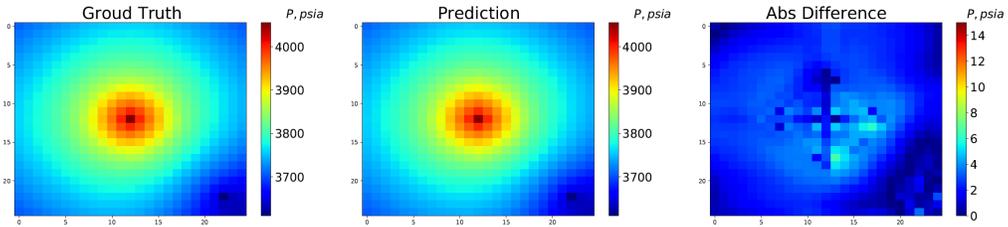

(b) Pressure from GDNN-Option 1: mean absolute error = 2.363 psia, mean relative error = 0.532%

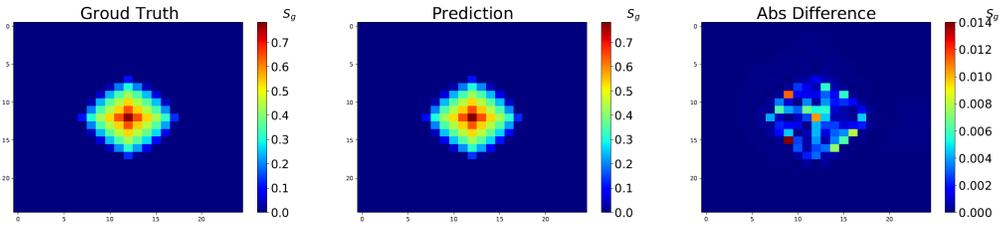

(c) Saturation from DNN: mean absolute error = 3.990e-4, mean relative error = 0.0510%

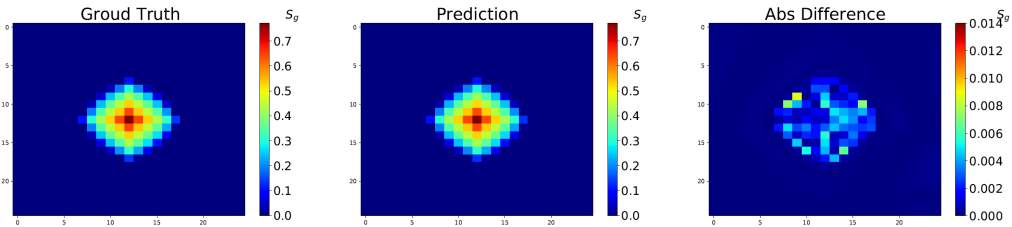

(d) Saturation from GDNN-Option 1: mean absolute error = 3.869e-4, mean relative error = 0.0500%

Fig. 7. Case 4 pressure and saturation predictions at 5.9 years from DNN and GDNN-Option 1 versus ground truth from numerical reservoir simulation. Left: ground truth of simulation results; middle: prediction from DNN or GDNN; right: absolute difference error.

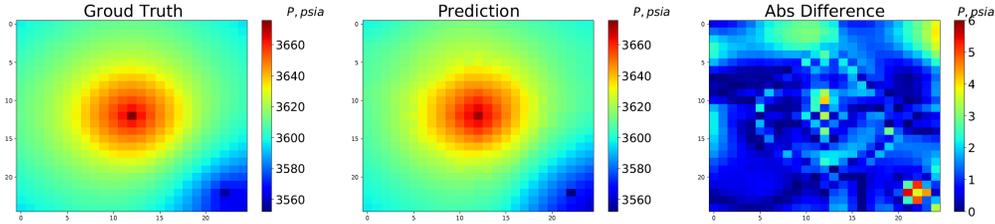

(a) Pressure from DNN: mean absolute error = 1.139 psia, mean relative error = 0.922%

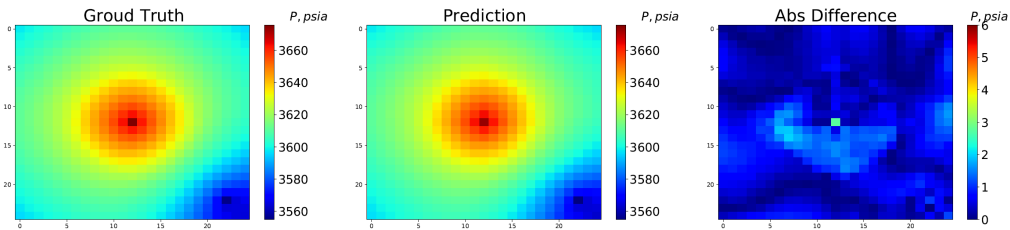

(b) Pressure GDNN-Option 1: mean absolute error = 0.526 psia, mean relative error = 0.435%

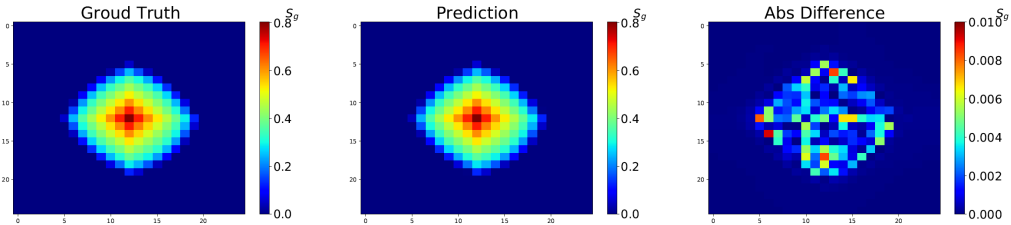

(c) Saturation DNN: mean absolute error = 6.317e-4, mean relative error = 0.0784%

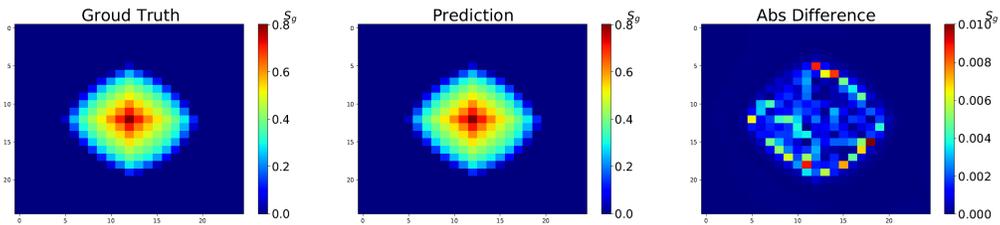

(d) Saturation GDNN-Option 1: mean absolute error = 5.430e-4, mean relative error = 0.0677%

Fig. 8. Case 14 pressure and saturation predictions at 5.9 years from DNN and GDNN-Option 1 versus ground truth from numerical reservoir simulation. Left: ground truth of simulation results; middle: prediction from DNN or GDNN; right: absolute difference error.

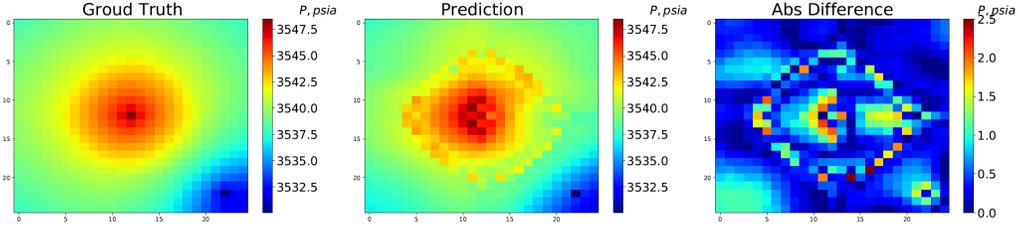

(a) Pressure from DNN: mean absolute error = 0.498 psia, mean relative error = 2.700%

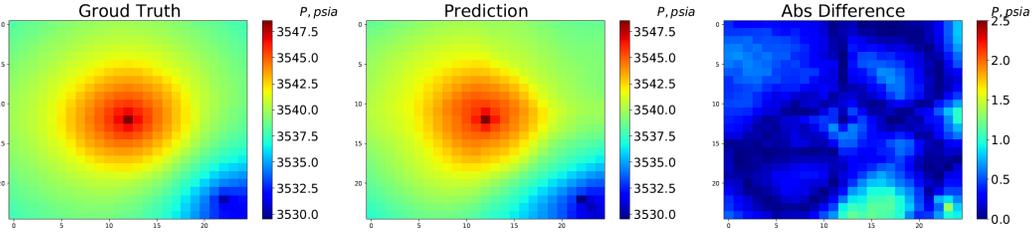

(b) Pressure from GDNN-Option 1: mean absolute error = 0.326 psia, mean relative error = 1.715%

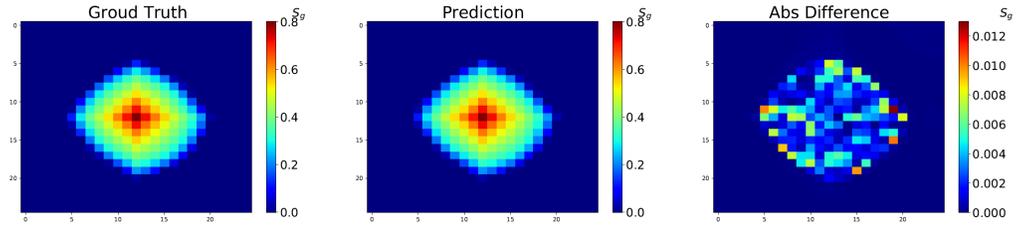

(c) Saturation from DNN: mean absolute error = 8.500e-4, mean relative error = 0.106%

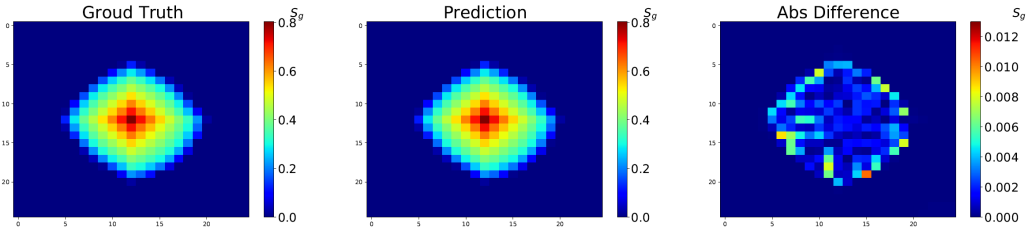

(d) Saturation from GDNN-Option 1: mean absolute error = 6.209e-4, mean relative error = 0.0772%

Fig. 9. Case 25 pressure and saturation predictions at 5.9 years from DNN and GDNN-Option 1 versus ground truth from numerical reservoir simulation. Left: ground truth of simulation results; middle: prediction from DNN or GDNN; right: absolute difference error.

### 3.2 Heterogeneous Three Dimensional Two-Phase Flow

In this section, we use a 3D heterogeneous reservoir model with 25 by 25 by 3 grid cells in the $x$, $y$ and $z$ directions respectively. The grid dimension is 300 $ft$ by 300 $ft$ by 11.1 $ft$ in the $x, y$ and $z$ directions, respectively. As shown in **Fig. 10**, the permeability and porosity fields are heterogeneous. The locations of the $CO_2$ injection well and the water production well is similar to that of the 2D case, and both wells are perforated across all 3 layers of the reservoir. The $CO_2$ injection well is operated with a maximum bottom pressure of 5000 psia and a maximum injection rate ranging from 1000 to 9000 $mscf/day$. The water production well is set to a constant bottom-hole pressure of 3525 psia. The two-phase relative permeability values are calculated through a look-up table based on **Fig. 4**. The final simulation time is 5.7 years with 68 time steps. There are 18 simulation runs with multiple combinations of permeability and porosity fields and the $CO_2$ injection rate, as shown in **Table 3**, including 2 pairs of permeability and porosity realizations and 9 $CO_2$ injection rates. Here the split of training and testing samples for deep learning is based on the three parameters in **Table 3**, including 16 cases (88% of samples) for training and 2 cases (12% of samples) for testing.

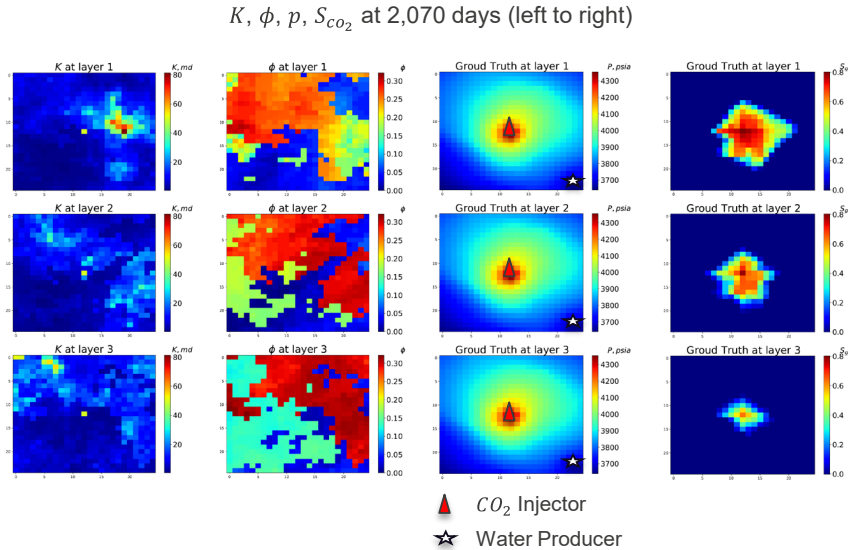

$K, \phi, p, S_{CO_2}$ at 2,070 days (left to right)

▲ $CO_2$ Injector
☆ Water Producer

Fig. 10. 3D reservoir model. Column 1: permeability of 3 layers; column 2: porosity of 3 layers; column 3: pressure snapshots of each layer; column 4: $CO_2$ saturation snapshots of each layer.

Table 3. Sampling parameters for 3D Reservoir Model

| Parameters | Number of Samples | Values |
| --- | --- | --- |
| Permeability $K$ | 2 | P50, P90 |
| Porosity $\phi$ | 2 | P50, P90 |
| $CO_2$ injection rate, $mscf/day$ | 9 | 1000 to 9000 |

Since the permeability is heterogeneous in this case, the spatial differential operators for permeability are included in the feature list (refer to Equation (10)). Further, the GDNN-Option 2 in **Table 1** is adopted with a fluid flux term regularizing the training loss function. Since the total number of samples for this problem reaches up to 2 million, the samples are subdivided into mini-batches of 50,000 samples/batch, and trained with an Adam optimizer. We used a grid searching analysis to determine the architecture of the deep learning models, including the number of neurons per hidden layer and the number of hidden layers. Through grid searching, the optimal number of neurons per hidden layers is 50 and the optimal number of hidden layers is 9. Since during training, the predicted values of pressures and saturations can become negative, a smooth softplus function $f(x) = ln(1 + e^x)$ is imposed at the output layer [23] to ensure positiveness.

It is worth mentioning that in order to capture nonlinearities, an adaptive activation function is implemented with PyTorch based on [25] as,

$$x^k = \sigma\left(a\mathcal{L}_k(x^{k-1})\right) \tag{12}$$

Different from classical non-adaptive activation functions, there is a hyper-parameter $a$ in the activation function $\sigma$ controlling the gradient to avoid the common issue of gradients exploding or vanishing in the training backpropagation, and the parameter $a$ of each hidden layer is determined during the training process. Different adaptive activation functions are evaluated, and ultimately it was determined that the adaptive Tanh function $AD\_Tanh(a,x) = \frac{e^{ax}-e^{-ax}}{e^{ax}+e^{-ax}}$ produces the lowest loss function, as shown in **Fig. 11**. Therefore, we utilize the adaptive Tanh function in this problem.

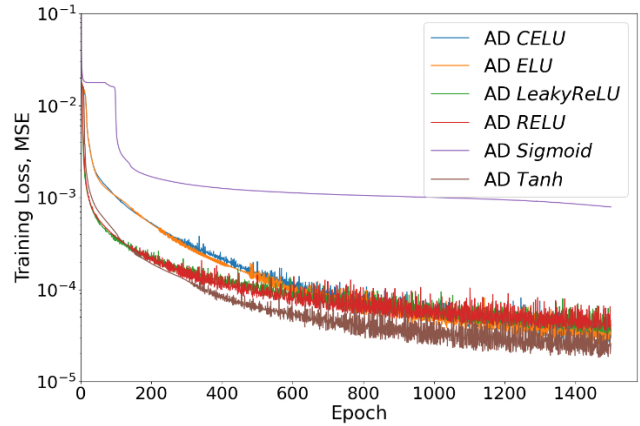

Fig. 11. Convergence of different adaptive transfer function

Both DNN and GDNN-Option 2 are evaluated using the architecture described above. Since DNN and GDNN-Option 2 have the same neural network architecture, GDNN-Option 2 uses a transfer learning approach by transferring the weights and bias from a trained DNN. In **Fig. 12**, we compare the temporal error propagation for pressure and saturation in DNN and GDNN-Option 2 models for the testing cases. The temporal errors for pressure for both models are fairly stable and the average error for GDNN-Option 2 is about 0.45% less than that for the DNN model. Similarly, both models have low saturation errors with maximum errors of 0.56% for DNN and 0.44% for GDNN-Option 2.

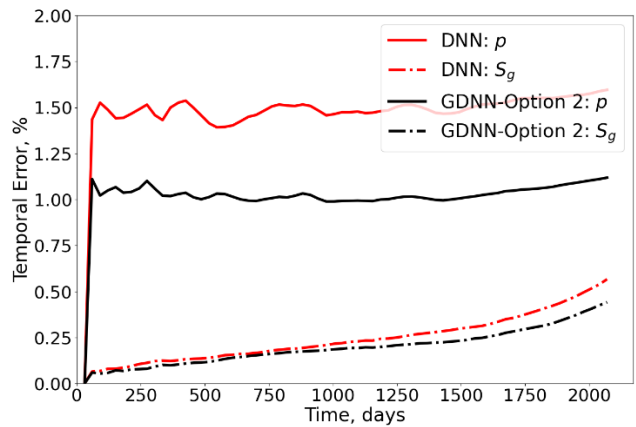

Fig. 12. Temporal error propagation for pressure and saturation in heterogeneous scenario

Next, the predictions of pressures and saturations for 2 different testing cases will be analyzed. **Table 4** lists the parameter values for the two testing cases. In the P90 realizations of permeability and porosity fields (Case 13), there are 13 inactive grid cells (with zero porosity) in the third layer of the reservoir which will result in discontinuous spatial derivatives for pressures that may lead to errors in predictions. Given that the temporal predictions are relatively stable (**Fig. 12**), we only compare snapshots of pressure and saturation at the time step 4.1 years.

Table 4. Value of parameters for 2 different testing cases

| Case | Permeability | Porosity | $CO_2$ injection rate, $mscf/day$ |
|------|--------------|----------|-----------------------------------|
| 6    | P50          | P50      | 4000                              |
| 13   | P90          | P90      | 3000                              |

The pressure and saturation predictions for Case 6 and Case 13 are shown in **Figs. 13** and **14,** respectively. For pressure, the DNN model can approximately capture the pressure plume shape, but its predicted pressure plume is not smooth particularly near the plume boundary. For the 2D homogeneous problem it was observed that DNN with similar architecture generates smooth pressure predictions including at the plume boundary, so it is likely that DNN by itself may not be able to effectively capture the irregular plume shapes of pressure and saturation resulting from the heterogeneous permeability and porosity distribution. For GDNN-Option 2, the mean relative error of pressure decreases by about 0.47% (3.8 psia) for Cases 6 and 13 compared to DNN, and the pressure prediction of GDNN-Option 2 becomes much smoother than that of DNN. As for the saturation, both DNN and GDNN-Option 2 accurately predict the irregular $CO_2$ plume shape caused by permeability and porosity heterogeneity with mean absolute error less than 1.0e-3.

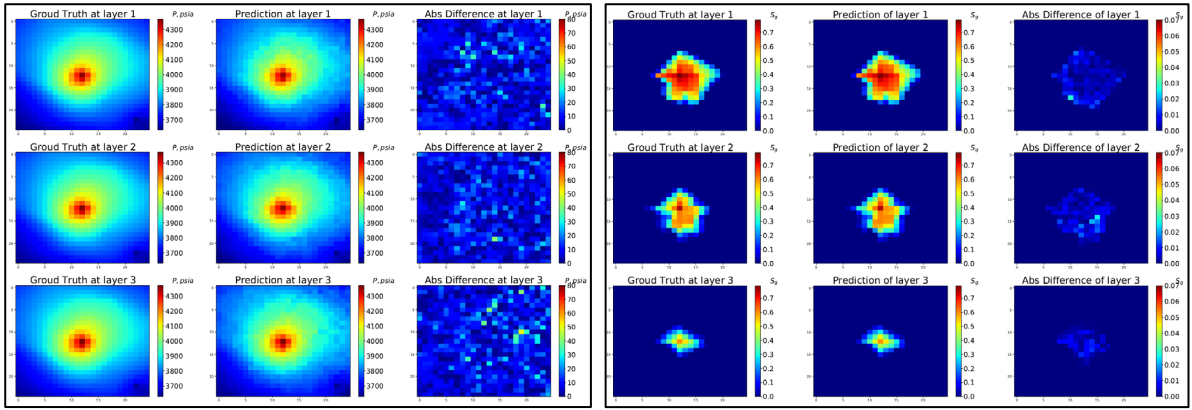

(a) DNN. Left – pressure: mean absolute error = 8.755 psia, mean relative error = 1.184%; right – saturation: mean absolute error = 7.907e-4, mean relative error = 0.1002%.

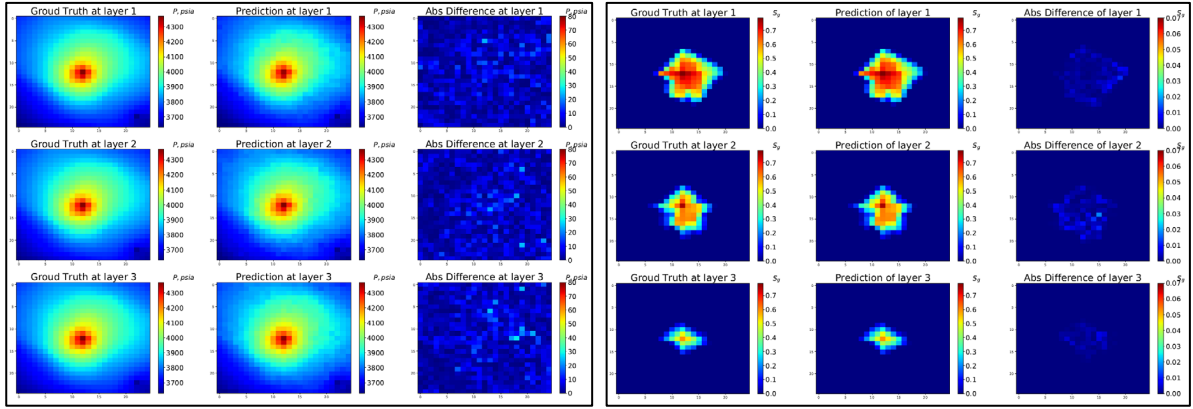

(b) GDNN-Option 2. Left – pressure: mean absolute error = 4.602 psia, mean relative error = 0.6221%; right – saturation: mean absolute error = 4.130e-4, mean relative error = 0.05235%.

Fig. 13. Pressure and saturation prediction of Case 6 of all the 3 layers at 4.1 years

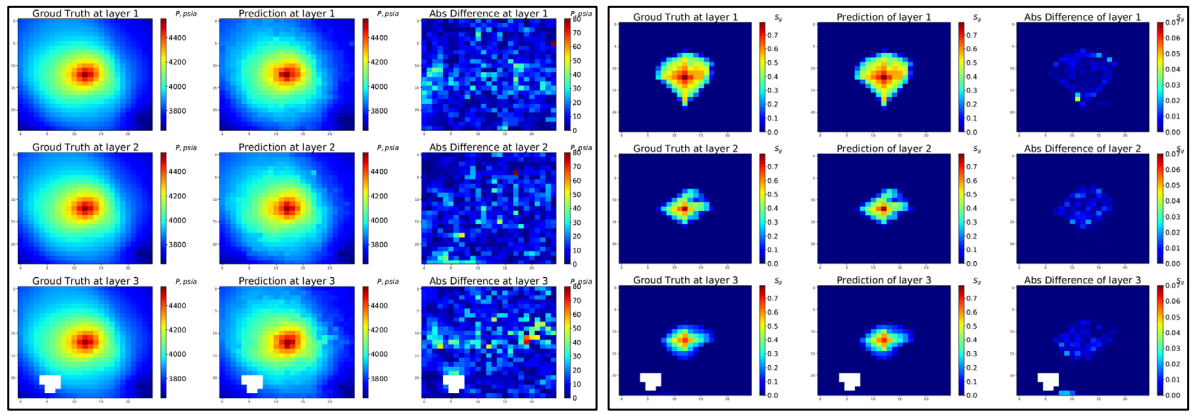

(a) DNN. Left – pressure: mean absolute error = 10.38 psia, mean relative error = 1.136%; right – saturation: mean absolute error = 8.067e-4, mean relative error = 0.1019%.

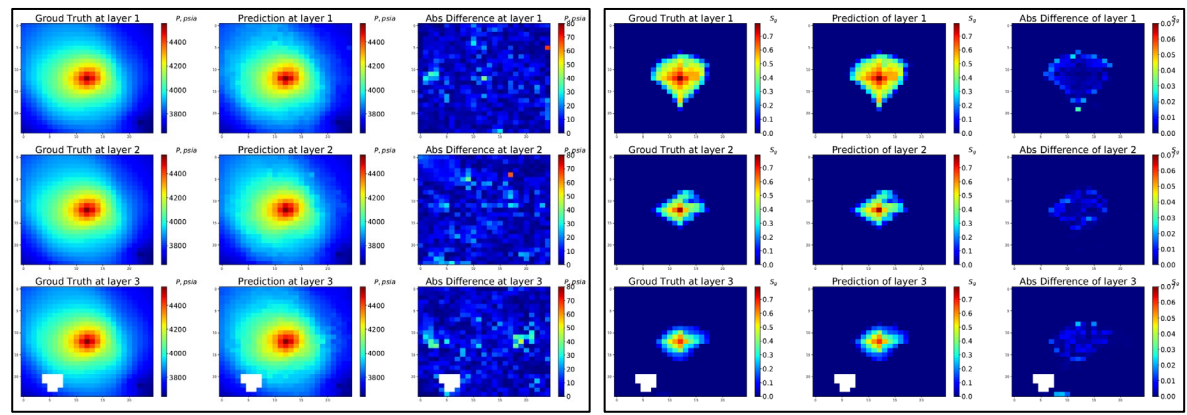

(b) GDNN-Option 2. Left – pressure: mean absolute error = 6.907 psia, mean relative error = 0.7555%; right – saturation: mean absolute error = 8.287e-4, mean relative error = 0.1046 %.

Fig. 14. Pressure and saturation prediction of Case 13 of all the 3 layers at 4.1 years

The CPU time at the training and epoch levels for the heterogeneous scenario are compared in **Table 5**. For consistent performance comparison, we also ran a GDNN-Option 1 (**Table 1**) with transfer learning from the trained DNN. The training time of GDNN takes into account the DNN training time because of transfer learning. It shows at the epoch level without I/O overhead DNN is 1.7 times faster than that of GDNN-Option 1, and is 4.3 times faster than that of GDNN-Option 2. Even though it is not the focus of this work, the training performance of GDNN can be improved in the future for multi-GPU computing or other alternative fast differential operator approximation approaches [39].

Table 5. Training Cost Comparison for 3D Heterogeneous Case

| Model | Training Time, (min) | Epoch Time, (sec) |
|---|---|---|
| DNN | 68 | 1.37 |
| GDNN-Option 1 | 257 | 3.78 |
| GDNN-Option 2 | 432 | 7.28 |

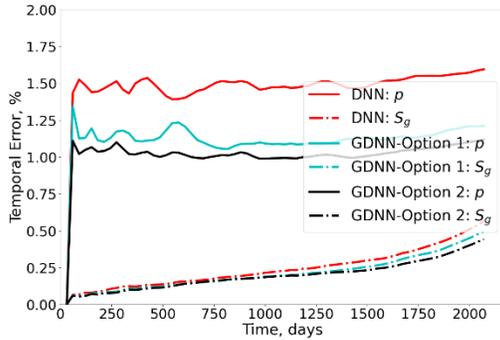

(a) Temporal error propagation

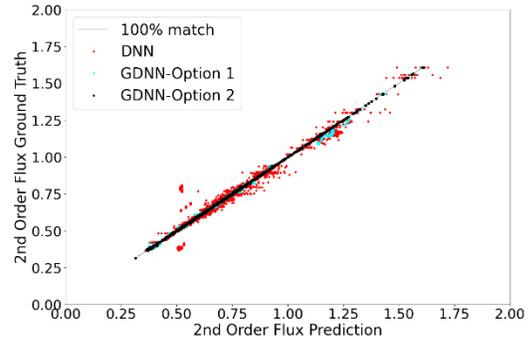

(b) Cross-plot of second order flux $\nabla \cdot J$

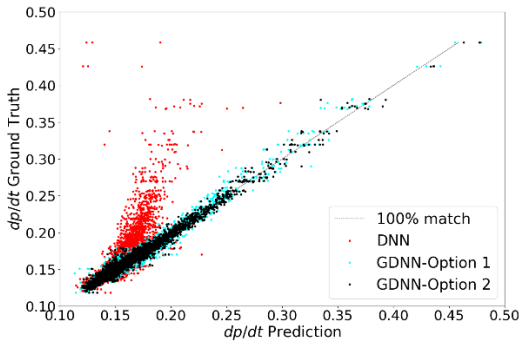

(c) Cross-plot of $\frac{\partial p}{\partial t}$

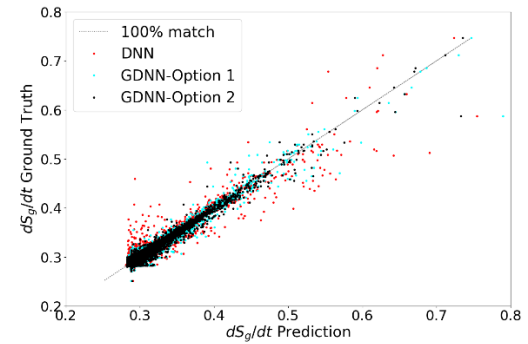

(d) Cross-plot of $\frac{\partial S_g}{\partial t}$

Fig. 15. Predictive metrics comparison of different models based on testing data

For the heterogeneous scenario, we also analyzed the accuracy of different loss sub-terms in DNN and GDNN options based on the testing results. The differential operators used in the loss function for GDNN-

Option 2 result in gradual decrease of the temporal errors of pressure and saturation (**Fig. 15(a)**). In **Fig. 15(b)**, the second order flux approximation $\nabla \cdot J$ has higher scatter for DNN compared to the GDNN-Option-1 while the GDNN-Option 2 ($R^2$ score 0.9999) has the least scatter. This result indicates that DNN by itself cannot learn the information carried by the second order flux terms just by minimizing the mismatch in state variables during the training process. In **Fig. 15(c)**, DNN severely underestimates the pressure derivative with respect to time ($R^2$ score 0.4463) which can help explain the worse pressure temporal error propagation for DNN observed in **Fig. 15(a)**. Both GDNN options have the same regularization terms related to $\frac{\partial p}{\partial t}$ and $\frac{\partial S_g}{\partial t}$ (**Table 1**) and thus their accuracy for the two terms is close to each other (**Fig. 15(c) and (d)**).

## 4. Conclusions

We have developed a GDNN model to predict multiphase flow and resulting reservoir changes (pressures and saturations). Specifically, augmented differential operators are used as first principles prior knowledge to regularize the training of the GDNN model. We successfully apply the methodology in the modeling of $CO_2$ injection into saline aquifer formations where brine is produced to manage reservoir pressure, and use it to predict subsurface responses such as pressure and supercritical phase saturation. We applied GDNN to a homogeneous case and a heterogeneous case. Our use of first order differential operators in the loss function used for GDNN (GDNN Option 1) improves predictions of pressure and supercritical phase saturation with good stability and smoothness. For the heterogeneous case, the time-space dependent evolution of pressure is impacted by the advective flux, and addition of second order flux terms based on pressure and permeability derivatives into the loss function significantly enhances the smoothness of pressure prediction. We also observed that saturation is easier to predict compared to pressure using either DNN or GDNN. This is primarily because the saturation front moves more slowly than the pressure front and thus does not experience similar dynamics or effect of permeability heterogeneity.

**CRediT authorship contribution statement**

**Bicheng Yan**: Study conceptualization, methodology development, software development, computational analysis, manuscript preparation and revision. **Dylan Robert Harp**: Supervision, study conceptualization, methodology development and manuscript revision. **Rajesh Pawar**: Funding support, supervision, study conceptualization, methodology development and manuscript revision.

**Acknowledgements**

The authors acknowledge the financial support by the US DOE through the Science-informed Machine Learning to Accelerate Real Time Decisions in Subsurface Applications (SMART) project. The SMART project is funded by US DOE Fossil Energy's Program Office's Carbon Storage Program and is managed by the National Energy Technology Laboratory (NETL). The authors also thank Dr. Seyyed A. Hosseini from University of Texas at Austin for providing the reservoir simulation data used for model development, training and testing and Dr. Diana Bacon from Pacific Northwest National Laboratory for providing a parsing tool to process the simulation data.